\documentstyle[prb,aps]{revtex}
\begin{document}
\draft
\title{Whither Correlated Electron Theory? }
\author{A. J. Millis}
\address{Center for Materials Theory\\
Department of Physics \& Astronomy, Rutgers University\\
136 Frelinghuysen Road, Piscataway, NJ 08854}
\date{\today}
\maketitle

\begin{abstract}
This is the text of the 'Theory' opening talk at the 2001 Strongly
Correlated Electron Systems conference. It contains opinions about some of
the outstanding scientific challenges facing the theory side of the
correlated electrons field.
\end{abstract}

\pacs{}

\section{\protect\bigskip Introduction}

The intellectual excitement of the 'Strongly Correlated Electrons' field
comes from the amazing variety of behaviors exhibited by actual materials;
behaviors which apparently cannot be understoond within the well established
framework of electronic condensed matter physics, and which suggest the
existence of as yet undiscovered classes of collective behavior. I think the
organizers of the 2001 Strongly Correlated Electron Systems (SCES)
conference have done a wonderful job of selecting talks which exhibit the
diverse and fascinating behavior of correlated systems. In this, the
'Conference Introduction: Theory' talk , I will present my own
opinions of the intellectual challenges and opportunities presented to the
theoretical side of our field. I am sure that this audience will disagree
with many aspects of what I have to say, and I\ look forward to hearing why.

I would like to begin by explaining what I mean by 'the well established
framework of \ electronic condensed matter physics', and then to go on by
building up a series of ways in which that framework has recently been
challenged or extended. I shall be concerned mainly with the excitations
which govern transport and other response functions and will not discuss
energy calculations at all. The conventional picture is based on Landau's
fermi liquid theory \cite{Pines63}, which shows how in many circumstances
the low energy physics of interacting electrons in solids can be reduced to
the physics of basically free electrons, moving in self consistent fields
generated by other electrons. The self consistent fields may lead to a
condensation into a paired state (either of particle-hole pairs, giving rise to
a magnetic or charge density wave state, or of  particle-particle pairs,
giving rise to a superconducting state). The leading corrections to this
picture involve electron-electron, electron-phonon, or electron-collective
mode scattering and may be treated basically perturbatively, for the usual
phase space reasons. In more formal language, these calculational techniques
and understandings are assoicated with the existence of an infrared stable
(or marginally unstable) fixed point (the fermi liquid fixed point discussed
by Shankar \cite{Shankar94}) around which one can expand with confidence.

This set of ideas (coupled with the tremendously successful complex of ideas
and techniques associated with density functional band theory) describes
many materials quite well; our field is concerned with those where it fails.
One clear example is the quasi one dimensional materials. It is by now
theoretically well established that strictly one dimensional 
electron systems are described by a Luttinger liqud \cite{Haldane81}, 
rather than a fermi  liquid fixed point.  The
properties of Luttinger liquids are well understood, and while there are
serious and important questions related to the degree to which any actual
material with a strong unidirectional anisotropy is well described by a
Luttinger model, I will not discuss these questions here. Another very well
studied set of phenomena concerns the electron gas in a very high magnetic
field. In a certain sense this involves the artificial creation (by
suppression of kinetic energy) of a non-fermi-liquid. To keep this
article within reasonable bounds I will restrict attention to zero magnetic
field.

In what follows I discuss the following topics, not all of which are
entirely orthogonal:

\begin{itemize}
\item  The approach to the fermi liquid fixed point

\item  New kinds of ordered phases

\item  Problems with the present theory of quantum criticality in metals

\item  The possible existence of and characterization of 'Non-fermi-liquid'
phases

\item  Non-quasiparticle physics: the exciting new techniques for dealing
with $T>0$ and with the 'incoherent' part of the electron spectral function.

\item  The importance, demonstrated to us by the 'colossal'
magnetoresistance manganites, of first order phase transitions and of
regimes of strong fluctuation.
\end{itemize}

\section{Leading corrections to fermi liquid theory: the
Belitz-Kirkpatrick-Vojta $\left| q\right| $.}

Fermi liquid theory is now \cite{Shankar94} understood as an almost infrared
stable fixed point of the interacting electron problem (''almost'' because
fermi liquids are generically marginally unstable to superconductivity). The
location of the Fermi surface, the quasiparticle fermi velocity and residue
and the Landau interaction parameters are parameters characterizing the
fixed point. Most workers implicitly or explicitly expected that (with one
exception) corrections to physical quantities, for example the free energy or
thesmall momentum transfer ($q$) limit of the 
spin or charge susceptibility were analytic
in the variables $\left( T/T_{F}\right) ^{2}$, $\left( q/p_{F}\right) ^{2}$,
corresponding in a straightforward way to irrelevant operators of dimension $%
2$. These $\left( T/T_{F}\right) ^{2}$, $\left( q/p_{F}\right) ^{2}$ terms
constitute among other things the bare temperature dependence of parameters
in theories of quantum critical phenomena \cite{Millis93}; also, the
coefficient of the $\left( q/p_{F}\right) ^{2}$ term in the current-current
correlation function gives the Landau diagmagnetism. The one exception, a $%
T^{2}\ln T_{F}/T$ term ($T$ in $d=2$) term in the specific heat coefficient $%
\gamma =C/T$, was understood as a special effect involving the effect of
long wavelength collective modes on the specific heat only. Because it was
believed to have a coefficient entirely expressible in terms of Landau
parameters and the fermi velocity, it was thought of as a feature of the Fermi
liquid fixed point \cite{Pethick76}.

Very interesting recent work of Belitz, Kirkpatrick and Vojta \cite{Belitz95}
challenges this understanding. These authors showed, by examination of
leading nontrivial terms in a perturbation series, that the leading momentum
dependence of the spin susceptibility $\chi _{s}$ of a clean fermi liquid is 
\begin{eqnarray}
\hspace{0.11in}\hspace{0.11in}\chi _{s}(q,\omega &=&0)=\chi _{0}+\left|
q\right| \hspace{0.11in}(d=2) \\
\chi _{s}(q,\omega &=&0)=\chi _{0}+q^{2}\ln (p_{F}/q)\hspace{0.11in}(d=3)
\end{eqnarray}
In the same approximation the dependence of the charge susceptibility is
simply $q^{2}$. \ This result was later confirmed and clarified by a
numerical analysis of the same set of perturbative graphs \cite{Hirashima00}%
. A subsequent systematic analysis \cite{Chitov01a} of the (mathematically
much simpler) possibility of a $\left| T\right| $ $(d=2)$ temperature
dependence clarified some aspects of the connection to fermi liquid theory
and demonstrated the importance of $^{\prime }2p_{F}^{\prime }$ terms to the
phenomena.

The results obtained so far have however been essentially calculational. A
general understanding has apparently not been achieved, but would be
desirable. For example, if the Belitz et. al. $\left| q\right| $ occurred in
the current-current correlation function, it would correspond to a strongly
divergent Landau diamagnetism! While the logarithmic correction expected in $%
d=3$ may be difficult to unambiguosly observe, a detailed experimental study
of quasi two-dimensional materials (e.g. $Sr_{2}RuO_{4}$ and especially $%
Sr_{3}Ru_{2}O_{7},$ which is near a ferromagnetic instability \cite{327})
would be very interesting.

\section{Novel forms of long ranged order}

Strongly correlated materials tend to order, and the search for new forms of
order is a central theme in our field. Recent attention has focussed on two
possibilities: orbital order, and 'staggered flux' or 'd-density wave' order.

{\it Orbital order} has recently been extensively discussed in the context
of the 'CMR' manganites \cite{Tokurabook} and it, or its fluctuations, may
be important in many other systems as well \cite{V2O3,OrbheavyF}. The idea
is that if the electrically active (and correlated) ion (e.g. the $Mn$ in
CMR compounds) sits at a site of sufficiently high symmetry (as the $Mn$
site has cubic symmetry in the ideal perovskite structure), then several
different orbital states (e.g. the $Mn$ $d_{3z^{2}-r^{2}}$ and $%
d_{x^{2}-y^{2}}$ orbitals) may be degenerate, and a spontaneous breaking of
the orbital symmetry may occur, in close analogy to the spontaneous breaking
of spin symmetry in a magnet. Similarly, in analogy with the 'spin liquid'
states extensively discussed in the context of high-$T_{c}$ materials (see
below, also), 'orbital liquid' states have been considered \cite
{Nagaosa96,Khaliulin99}. Note, though, that orbital ordering generically
breaks a crystal symmetry; it thus tends to be strongly coupled to lattice
distortions, which should be ncluded in any theory and which tend to make
the situation less quantum mechanical. Further, orbital index (unlike spin
index) tends not to be conserved by the electron hopping part of the
Hamiltonian, rendering 'orbital liquid' states different from spin liquids.

{\it Staggered flux }or{\it \ d-density wave} ordering falls into the
familiar class of particle-hole pair condensation, but the order parameter
involves intersite correlations in a crucial way and in particular leads to
the formation of lattice scale circulating current patterns. The idea was
introduced in the context of high temperature superconductivity by Affleck
and Marston \cite{Affleck91}. In their theory the circulating currents were
a consequence of a particular kind of spin correlations produced (in the
approximation used in Ref \cite{Affleck91}) by the competition between
magnetic ordering and carrier motion. More recently, Chakravarty et al \cite
{ddw} have reintroduced the concept (still in the context of high-$T_{c}$)
and argued that one should treat the circulating currents as fundamental,
and Chandra and co-workers \cite{Chandra01} have proposed that something of this kind is the
mysterious hidden order parameter of $URu_{2}Si_2$

\section{Quantum Criticality in Metals}

An obvious place to seek for the breakdown of conventional concepts is in
the vicinity of a second order $T=0$ phase transition (quantum critical
point), where the diverging correlation length suggests the existence of a
long ranged, unscreened interaction which would invalidate the usual
justification for fermi liquid theory. The theory of quantum criticality in
metals was founded by Hertz \cite{Hertz76} and studied in more detail by the
author \cite{Millis93} and many others. The basic starting point is an
action $S$ for an order parameter $\phi $, obtained by formally integrating
out the electron excitations and of the form 
\begin{equation}
S=S_{dyn}+\int d^{d}xd\tau \left[ \xi _{0}^{2}\left( \nabla \phi \right)
^{2}+r\phi ^{2}+u\phi ^{4}+...\right]
\end{equation}
where $S_{dyn}$ expresses the dynamics of the order parameter. In metals,
the crucial point is that the excitations are overdamped, ($S_{dyn}\sim
i\left| \omega \right| /\Gamma _{q}$ with $\Gamma _{q}$ a damping
ocefficient) leading to a dynamical exponent $z\geq 2$ \cite{Millis93}. The
effecitve dimensionality is $d_{eff}=d+z$ and for $d\geq 2$ $d_{eff}\geq 4$
so one expects the transition to be at or above its upper critical dimension
and described by a Gaussian model with a dangerously irrelevant operator \cite
{Millis93}. The theory is then very well controlled and leads to a number of
predictions which agree roughly with experiment in some cases \cite
{ZrZn2,CePd3} but in many cases disagree strongly \cite{CeCu6,aniso}. The
reason for the disagreement is presently unclear. In some cases it may be
associated with quasi-two-dimensional spin fluctuations \cite{quasitwod},
but I am uncomfortable with this explanation because one would expect a
crossover to three dimensional behavior to occur within the experimental
range. The detailed crossover behavior has however not been carefully
studied theoretically or experimentally, so it is possible that the numbers
are such that my concern is unwarranted.

Two other more theoretically interesting possibilities have been raised. One
is that the presence of gapless electron excitations can change the form of $%
S$. Chubukov and co-workers \cite{Chubukov01} have demonstrated that in a
model of two dimensional metallic antiferromagnetism the coefficient $u$ in
fact is not a constant but has a nontrivial $\omega ,q$ dependence, leading
to somewhat different behavior for response functions etc. In general one
ought uncomfortable with integrating out gapless excitations, so it seems
that this issue deserves examination in other contexts as well.

A\ more radical proposal was made by Schroeder, Coleman and co-workers 
\cite{Schroeder99} in the context of analysing data on the $Ceu_{6-x}Au_{x}$
system, which undergoes a paramagnet-antiferromagnet transition at
approximately $x=0.1$ with critical properties (most notably the linear
scaling of the Neel temperature with control parameter and a logarithmically
divergent specific heat coefficient) strikingly inconsistent with the
standard analysis. These authors argued on the basis of scaling anayses of
the $\omega ,q$ dependence of neutron scattering measurements of the
susceptibility at a range of high momenta, along with the temperature and
field dependence of the uniform susceptibility, that at the critical point $%
S_{dyn}\sim \left( i\omega \right) ^{\alpha }$ with $\alpha \approx 0.75<1$.
In other words, at criticality, the very low frequency spin dynamics are
anomalous everywhere in the zone, not just in the vicinity of the ordering
wavevector. These authors speculated that this was a signature of a new kind
of critical point, at which the Kondo temperature or renormalized fermi
energy vanished at precisely the point at which magnetism began \cite
{Schroeder99}; see also \cite{Si99}. Si and collaborators\cite{Si01}, 
suggestions of Sengupta \cite{Sengupta98} and Smith and Si \cite{Si97}
have found a model of
fermions coupled to two-dimensional bosons, which in the dynamical mean
field approximation appears to yield the sort of behavior suggested by
Coleman et. al. Further experimental and theoretical study of these
extremely interesting models and phenomena would be very desirable. One
issue which deserves clarification is the behavior of the specific heat
coefficient. Naively, a divergence of the local spin susceptibility implies
a similar divergence of the specific heat coefficient. The connection is not
completely rigorous, but from the Kubo formula one has 
\begin{eqnarray}
\sum_{q}\chi _{q}^{^{\prime \prime }}(\omega ) &=&\sum_{n}\left|
<0|m_{loc}|n>\right| \delta (\omega -E_{n}) \\
&=&\int d\varepsilon {\cal D}(\varepsilon )m_{loc}^{2}(\varepsilon )\delta
(\omega -E_{n})
\end{eqnarray}
where ${\cal D}(\varepsilon )$ is the density of states (of exact
eigenstates) and $m_{loc}$ is the matrix element by which a local magnetic
field couples to these eignestates. The proposed behavior of $\chi $ would
imply that either ${\cal D}(\varepsilon )/\varepsilon $ or $%
m_{loc}(\varepsilon )$ diverges as $\varepsilon \rightarrow 0$. As it seems
unlikely that a local matrix element would diverge, one is tempted to
ascribe the divergence to ${\cal D}(\varepsilon )/\varepsilon $, which would
imply a stronger than logarithmic divergence of the specific heat, in
contrast to observation on $CeCu_{6-x}Au_x$. Interestingly, Trovarelli et. al.
\cite{Trovarelli00} have observed that in $YbRh_2Si_2$ that an apparent
logarithmic divergence in the specific heat coefficient , occurring
for temperatures in the range of 1-5$K$ is changed at very
low $T$ to a stronger divergence.  

\section{Non Fermi Liquid Phases}

A long-standing goal of our field is the discovery of metallic phase with no
(obvious) long ranged order and with low temperature properties inconsistent
with the predictions of Fermi Liquid theory. Recent detailed high-precision
experimental studies have produced an intriguing finding. In a number of
heavy fermion and transition metal compounds exhibiting a pressure-tuned
magnetic critical point, on the paramagnetic side the low temperature
resistivity does not take the $T^{2}$ form expected from fermi liquid
theory, but instead exhibit a lower exponent, even down to the lowest
temperatures \cite{Lonzarich01}. It is possible that this is a crossover
phenomenon, but one should be able to estimate the crossover scale from the
behavior of the Neel temperature on the low pressure side of the transition,
and the temperatures attained are far below this. If these results hold up,
they present a fundamental challenge to our understanding of metal physics.

On the theory side, note that the conventional dichotomy in the 'Kondo
lattice' model describing heavy fermion materials is that one either
dissolves the local moments into the conduction electrons, forming a heavy
fermi liquid, or one magnetically orders them \cite{Doniach78}. Kikoin and
co-workers \cite{Kikoin97} have argued that a third alternative is possible,
namely to form a 'spin liquid' state from the local moments, leaving the
conduction electrons to form a small fermi surface. My understanding is that
experiments so far do not support this possibility, but it is a very
interesting theoretical suggestion which seems very much worth further
investigation. Senthil, Fisher and co-workers \cite{Senthil01} have argued
that the theoretical search for new physics should focus on 'fractionalized'
phases in which at least some of the excitations exhibit novel quantum
numbers (e.g. charge 0 and spin 1/2)\cite{explain}, and (following work by
Wen in the quantum hall context) have shown that this implies the presence
of a topological order revealed most clearly by the ground state degeneracy
in a multiply connected sample. Luillier and co-workers have presented very
suggestive numerical evidence that spin liquid phases (which, if possessing
spin-Peierls order, do so on such a long length scale as to leave a
reasonable regime of spin linquid behavior) actually occur in certain two
dimensional frustrated Heisenberg spin models \cite{Lhuillier00}, and Sondhi
and Moessner \cite{Moessner00} have convincingly demonstrated the existence
of a spin liquid in the (perhaps less directly physical) tirangular-lattice
quantum dimer model.

\section{Beyond particles}

Much attention has focussed on properties controlled either by ground state
symmetries or by the excitation of a small number of quanta ('particles') in
part because the theoretical treatment is simplest and the differents
between phases perhaps least unambiguous. In the fermi liquid context one
focusses on properties controlled by the quasiparticle pole of the Green
function (or by the long-wvelength collective modes); in the
non-fermi-liquid context one searches for new phases, and new particles
(e.g. 'semions' or 'spinons'). However, many situations arise in which a
description in terms of a small number of (reasonably well-defined)
particles is not appropriate, and I would like to draw your attention to
recent very interesting progress in this area.

Conventionally, one thinks of superconductivity as arising from the pairing
of two reasonably well-defined particles (although Eliashberg theory allows
one to go a bit beyond this assumption). Chubukov and co-workers studied
superconductivity near a two dimensional antiferromagnetic critical point 
\cite{Chubukov01b}, where the electron Green function is very broad and has
no coherent part at all, and were able to formulate a theory of 'incoherent
pairing'. I think this work is important in a broader context, as an example
of physics without (quasi)particles, and could be relevant to the
long-standing mystery of $UBe_{13}$\cite{UBe13}$,$ where superconductivity
apparently emerges from an extremely highly resistive and strongly scattered
state.

A characteristic feature of a strongly incoherent state is a lack of strong
momentum dependence (put differently, the crucial aspect of a particle is
its energy-momentum dispersion relation, which is absent or weak in a
strongly incoherent state). This feature allowed Chubukov et. al. to make
progress by reducing the problem to a self-consistent set of integral
equations involving only frequency. The consequence of negligible momentum
dependence (in this case of the electron self energy) has been exploited to
a much greater extent in the 'dynamical mean field' or '$d=\infty ^{\prime }$
approach to correlated electron problems, introduced by Mueller-Hartmann,
pursued by many workers, and brought into its present highly successful and
useful form by Kotliar and many co-workers \cite{dmft}. This line of
approach has had many successes, including the elucidation of many key
features of the Mott metal-insulator transition and its recent 'marriage' to
LMTO band theory (enabling the more or less ab-initio calculation of key
features of the phase diagram of plutonium and of the magnetic transition
temperatures of $Fe$ and $Ni$). Here, however, I would stress the great 
{\it conceptual} importance of the method. The essential idea is that if the
momentum dependence of the electron self energy is negligible, then all
important physical quantities may be derived from a functional of the local
green function $G_{loc}$ given by 
\begin{equation}
G_{loc}(\omega )=\int \frac{d^{d}p}{\left( 2\pi \right) ^{d}}\frac{1}{\omega
-\varepsilon _{p}-\Sigma (\omega )}
\end{equation}
where $\varepsilon _{p}$ is the underlying band dispersion. The key point
(due to Georges \ and Kotliar \cite{Georges93}) is that because $G_{loc}$ is
a function only of frequency, the functional of $G_{loc}$ may be written as
a quantum impurity and explicitly constructed via (not too heavy) numerics.
The important object in the theory, thus, is not a particle or field with a
well defined $\omega $ vs $k$ dispersion relation. Rather, it is a {\it %
function of frequency }which includes coherent and incoherent parts of
propagators, and $T>0$ and $T=0$ behaviors on the same footing. This
produces the ability to calculate many things which could not revious have
been calculated, and mor importantly, I believe will lead to new ways of
thinking about correlated electron physics.

A further advantage of the DMFT work is that there are a number of important
questions in correlated electron physics which are quantitative in nature,
yet extremely interesting, and which DMFT may help us
to understand. One topic of current discussion is 'Why is the
ferromagnetic transition temperature of lightly doped hexaborides so high?
Why does it drop rapidly for dopings beyond a few percent?'\cite{Hexaboride}%
. Another topic of long standing interest is 'Why is the pairing scale in
the hole-doped cuprates so high?'. This question has recently been joined by
the question 'Why is the pairing scale in hole-doped $C_{60}$ so high?'\cite
{Schoen01}

\section{First Order Transitions Lead to Interesting Behavior Also!}

A commonly held belief in correlated electron physics is that first order
phase transitions are not interesting, because they are not accompanied by a
diverging length. I wish to argue that this point of view is
perhaps overly narrow. First, it rules out of consideration the many
phenomena assoicated with metal insulator transitions (which are typically
first order with a second-order end point) and mixed valence phenomena (also
typically first order). \ Second, as is well known, correlated electron
materials generically have very low energy scales and several competing
phases; critical end-poionts of first order transitions may occur at much
lower temperatures than one would guess and as a result fluctuations may be
much stronger. Indeed, it is even possible by tuning a parameter to bring
the critical end point of a first order line is sufficiently close to zero
temperature that the physics is controlled by a {\it quantum critical end
point} \cite{Millis01c}. Apparently, the metamagnetic transitions in many
heavy fermion compounds are in this category (or may be made so by applying
e.g. pressure), and a very exciting recent development is that at ambient
pressure the nearly two-dimensional compound $Sr_{3}Ru_{2}O_{7}$ may be in
this category \cite{327}. Further, the classic Imry-Wortis work \cite{Imry76}
shows that in two spatial dimensions, randomness which couples to the energy
density (i.e. any randomness at all) converts first order transitions into
second order ones, leading (presumably) to a host of unusual phenomena. I
sometimes wonder if this phenomenon is at the root of the
difficulties involved in interpreting data on high-T$_{c}$ materials.

These ideas are very nicely exemplified by the results obtained in the last
five years on the 'colossal' magnetoresistance manganites. These
materials display an
amazing sensitivity of properties to external perturbations (and sample form
and quality!), of which the highly-touted 'colossal' \cite{Jin94}
magnetoresistance is
but one example. It is now generally accepted that linear response
susceptibilities are not particularly large, and that the exotic behavior is
due to a multiphase coexistence \cite{Cheong98} which is itself a
consequence of the presence of several competing phases (different varieties
of charge ordered insulater as well as ferromagnetic metal) with first order
phase boundaries between them, along with disorder, which leads to the
multiphase coexistence. Two particularly striking recent results are that in
the material $La_{0.7}Ca_{0.3}MnO_{3}$ (which has a ferromagnetic ground
state and a magnetic transition at $T_{c}\approx 260K$) the phase transition
is in fact first order \cite{Lynn00} and the high temperature phase exhibits
surprisingly strong charge/orbital fluctuations extending up to quite high
temperatures \cite{Lynn00,Kiryukhin01}. The percolation idea was picked up
by theorists, and modelled via numerical simulations of the two dimensional
random field Ising model \cite{DaGotto00}, with quite plausible agreement
with data. One difficulty is that the calculation relied on what the general
arguments \cite{Imry76} are quite special features of two dimensional
models, which would not be expected to hold in the three dimensional cases
of experimental relevance. It seems likely to the present author that the
relevant disorder is long-range correlated, and arises from strain fields.
This belief is supported by recent work showing a martensitic character to
the charge/orbital ordering transitions \cite{Cheong01}. In any event, the
example of the manganites shows that first order phase boundaries can lead
to fascinatingly complex behavior. It seems likely that this idea will be
fruitful in other areas of correlated electron physics as well.

\section{Summary}

In this article I have tried to present an overview
of important challenges and open issues in
correlated electron theory. I apologize to all those whose work I have
misrepresented or not represented.

\acknowledgements
My work in this area is supported by NSF DMR0081075 and the University of
Maryland-Rutgers MRSEC.

%%%%%%%%%%%%%%%%%%%%%%%%%%%%%%%%%%%%%%%%%%%%%%%%%%%%%%%%%%%%%%%%%%%%%%%
%%%%%%   REFERENCES   %%%%%%%%%%%%%%%%%%%%%%%%%%%%%%%%%%%%%%%%%%%%%%%%%
%%%%%%%%%%%%%%%%%%%%%%%%%%%%%%%%%%%%%%%%%%%%%%%%%%%%%%%%%%%%%%%%%%%%%%%


\begin{references}
\bibitem{explain}  For example, in fractionally quantized Hall states the
magnetoroton (a state of conventional quantum numbers) may lie lower than
the 'Laughlin quasiparticle'.
\end{references}
\end{document}